\documentclass[preprint]{elsarticle}

\usepackage{lineno,hyperref}
\modulolinenumbers[5]

\journal{Journal of \LaTeX\ Templates}

\usepackage{graphicx}
\usepackage{multirow}
\usepackage{enumerate}
\usepackage[shortlabels]{enumitem}
\usepackage{booktabs} % To add midrule and top rule in tables
\begin{document}

\begin{frontmatter}

%\title{A Community Strategy Framework: What to consider when determining where and how to obtain Influence on Requirements in Open Source Software Communities} %BR a bit long...

\title{A Community Strategy Framework --\\How to obtain Influence on Requirements in\\Meritocratic Open Source Software Communities?} %BR is this shorter version ok?

% \title{A Community Strategy Framework: \\ How to Gain Requirements Engineering Influence in Open Source Software Communities}

%Influence on Established Open Source Projects' Requirements: Where do you need it and how do you get it?}

% Framework, community
% Community Strategy Framework 

\author[LU]{J.~Lin{\aa}ker\corref{corrAuthor}}
\ead{johan.linaker@cs.lth.se}
\author[LU]{B.~Regnell}
\ead{bjorn.regnell@cs.lth.se}
\author[UVIC]{D.~Damian}
\ead{damian.daniela@gmail.com}

\cortext[corrAuthor]{Corresponding author}

\address[LU]{Lund University, Box 118, 221 00 Lund, Sweden}
\address[UVIC]{University of Victoria, PO Box 1700, STN CSC Victoria, BC V8W 2Y2, Canada}

\begin{abstract}
\textbf{Context:}
In the Requirements Engineering (RE) process of %BR: changed "in" to "of" ???
an Open Source Software (OSS) community, an involved firm is a stakeholder among many. Conflicting agendas may create miss-alignment with the firm's internal requirements strategy. In communities with meritocratic governance or with aspects %BR: removed of, added with
thereof, a firm has the opportunity to affect the RE process %BR: proposed changed: according to  -> in line with
in line with their own agenda by gaining influence through active and symbiotic engagements.
\textbf{Objective:}
The focus of this study has been to identify what aspects that firms should consider when they assess their need of influencing the RE process in an OSS community, as well as what engagement practices that should be considered in order to gain this influence.
\textbf{Method:}
Using a design science approach, 21 interviews with 18 industry professionals from 12 different software-intensive firms were conducted to explore, design and validate an artifact for the problem context.
\textbf{Results:}
A Community Strategy Framework (CSF) is presented to help firms create community strategies that describe if and why they need influence on the RE process in a specific (meritocratic) OSS community, and how the firm could gain it. The framework consists of aspects and engagement practices. The aspects help determine how important an OSS project and its community is from business and technical perspectives. A community perspective is used when considering the feasibility and potential in gaining influence. The engagement practices are intended as a tool-box for how a firm can engage with a community in order to build influence needed.
\textbf{Conclusion:}
It is concluded from interview-based validation that the proposed CSF may provide support for firms in creating and tailoring community strategies and help them to focus resources on communities that matter and gain the influence needed on their respective RE processes.

\end{abstract}


\end{frontmatter}

% \linenumbers

%==============================================
%==============================================
\section{Introduction}
\label{sec:Introduction}
%==============================================
%==============================================

Open Source Software (OSS) is for many firms today a fundamental building block for creating, delivering and supporting their product and service offerings, or internal operations~\cite{dahlander2008firms, butler2018investigation}. The development and maintenance of an OSS project are performed within a software ecosystem~\cite{jansen2013defining}, often referred to as a community. The members of a community consist of stakeholders of the OSS project, i.e., \textit{``\ldots person[s] or organization[s] who influences a system's requirements or who [are] impacted by that system''}~\cite{glinz2007guest}. In this case, ''a system'' refers to the OSS project. To a firm involved in an OSS community, the Requirements Engineering (RE) process in the community is an external process where the firm is no longer the central authority, in contrast to traditional market-driven RE~\cite{regnell2005market}. Instead, the firm is a stakeholder among many which may introduce conflicting agendas from other stakeholders~\cite{munir2015open, schaarschmidt2015firms, maenpaa2018organizing}, and a new type of power and politics than the firm might be used to~\cite{Milne2012}. Consequences may include a lack of control over what requirements that are implemented, and miss-alignment with the firm's internal RE process~\cite{dahlander2008firms, wnuk2012can}. A firm who wish to affect the RE process according to their agenda may, therefore, have to build up an influence within the community~\cite{schaarschmidt2015firms}.

With influence, we refer to the Merriam-Webster dictionary~\footnote{http://www.merriam-webster.com/dictionary/influence} which defines it as \textit{``the power to change or affect someone or something''}. In our context, this relates to the power of a firm to change or affect a requirement of interest in an OSS community, for example, how a requirement is specified, prioritized, and realized, both short-term in release-planning, and long-term on the road-map~\cite{german2003gnome, laurent2009lessons, noll2007innovation}. In OSS communities with a meritocratic governance structure~\cite{markus2007governance, de2007governance}, either in part or in full~\cite{shaikh2017governing}, influence is gained by proving merit and earning trust and status within the community~\cite{fielding1999shared}. What merit constitutes depends on the context~\cite{eckhardt2014merits, o2007emergence}, but is generally gained by building an active and symbiotic relationship with the community where a firm dedicates resources, contributes internal requirements and actively participates in the development of the OSS~\cite{dahlander2005relationships, dahlander2006man, butler2018investigation, schaarschmidt2015firms, syeed2017measuring, nguyen2018do}. A meritocratic OSS community, therefore, offers an opportunity for the focal firm to influence the community's RE process according to the firm's own agenda while competing and collaborating with the other stakeholders in the community~\cite{nguyen2018do}.

For a firm engaged in many communities, such investments may be costly if it is distributed over all communities. It may be that only a few communities are of such strategic importance to the firm, and are in a state where the firm needs to have an influence on their RE processes~\cite{dahlander2008firms}. For a strategic community that is healthy, predictable and aligned with a firm's internal agenda, it may be that a high level of influence is not motivated~\cite{butler2018investigation}. Therefore, to optimize its resource utilization and investments where best needed, firms may have to assess how they could benefit from a specific OSS project and its community, and then if and how much influence that is required to reap these benefits~\cite{butler2018investigation}. To the best of our knowledge, there is no systematic approach to perform this kind of assessment, why we pose our first research question as:

\begin{enumerate}[\textbf{RQ1}]
    \item What aspects should a firm consider when assessing its need to influence the RE process in a meritocratic OSS community?
\end{enumerate}

If a firm assesses that they need influence on the RE process in a meritocratic OSS community, the follow-up question is: what should their community engagement look like and how should they invest their resources to gain the influence needed? To the best of our knowledge, an overview on a software engineering level of what engagement practices that may be used to build influence in meritocratic OSS communities is absent (e.g.,~\cite{linaaker2015requirements, munir2015open, alves2017software, franco2017open}). This gap leads us to pose our second research question:

\begin{enumerate}[\textbf{RQ2}]
\item What practices should a firm consider to gain influence on the RE process in a meritocratic OSS community?
\end{enumerate}

To address these two research questions, this paper presents a Community Strategy Framework (CSF). A community strategy should describe if and why a firm needs influence on the RE process in a specific OSS community, and how the firm could gain it. Thus, the objective of CSF is to help firms create and tailor community strategies that enable them to focus resources on communities that matter and gain the influence needed on their respective RE processes. 

Using a design science approach~\cite{hevner2004design, wieringa2014design}, we leverage a series of ten semi-structured interviews with industry professionals to explore the problem context. Interview transcripts were then inductively coded~\cite{runeson2012casestudy} which resulted in a first design of the CSF. To validate and refine the design, seven interviews were conducted where the interviewees were presented with the CSF and asked questions regarding its completeness and correctness. To evaluate the applicability and utility of CSF~\cite{hevner2004design}, in one of these interviews, the framework was also applied to a fictitious example based on an earlier reported case study~\cite{munir2017open}. As the last step, a case validation was conducted by interviewing four industry professionals from a software-intensive firm engaged in multiple OSS communities. Questions focused on the validity of CSF in the context of the firm's community engagements. In total, we conducted 21 interviews with 18 industry professionals from 12 different software-intensive firms.

The rest of the paper is structured as follows: in Section~\ref{sec:RelatedWork}, we present related work, which this study builds upon. In Section~\ref{sec:ResearchDesign}, we present the research design of this study and how it was executed. In Section~\ref{sec:CommunityStrategyFramework}, we present the CSF, and in Section~\ref{sec:CaseExample} the framework is applied to a fictitious example. In Section~\ref{sec:Discussion} we discuss our findings, followed by a discussion on threats to validity in Section~\ref{sec:ThreatsToValidity}. In Section~\ref{sec:Conclusions}, we conclude the paper.

%==============================================
%==============================================
\section{Related Work}
\label{sec:RelatedWork}
%==============================================
%==============================================

In this section, we present the related work that provides a theoretical underpinning for the design of the artifact called the Community Strategy Framework (CSF). This theoretical basis is also used in the discussions on the validity of the proposed framework (see Section~\ref{sec:Discussion}). 
% First, we give a background on how RE in OSS communities may function and the need for influence to affect, e.g., selection and prioritization of requirements. Second, we present the role of governance and authority structure in an OSS community in regards to how influence may be gained. 
% Third, we continue on the role of governance and authority structure but focus on cases where a community is institutionalized through a foundation. 
% Third, we present earlier work on how firms can gain influence on the RE process in OSS communities, and fourth, we focus on previous work describing perspectives that may motivate the need for involvement in an OSS community. Lastly, we provide a summary and explain the research gap that this study aims to fill.

\subsection{Requirements Engineering in OSS communities}

Compared to classic RE~\cite{alspaugh2013ongoing}, OSS RE can be described as a collaborative, transparent and open process involving the stakeholders (both developers and users) in the community with interest in specific requirements~\cite{alspaugh2013ongoing, aagerfalk2008outsourcing}. Formal methods and processes, as well as documents or central repositories, are often absent~\cite{kuriakose2015how, castro2012differences}. Instead, a requirement may often be represented by multiple artifacts which are stored and managed in a series of interconnected and overlapping repositories, e.g., as an issue in an issue tracker and mail threads in a mailing list~\cite{scacchi2010collaboration}. These repositories also function as communication channels for the stakeholders where the requirements are asserted (i.e., elicited from the OSS community perspective), analyzed, and specified informally, and often realized simultaneously~\cite{bhowmik2018refinement, kuriakose2015how, ernst2012case, castro2012differences}. This is an iterative process characterized as just-in-time RE~\cite{ernst2012case, bhowmik2018refinement} and where the social interactions between the stakeholders are often decentralized and dynamic~\cite{bhowmik2015on}. However, these can on occasion also occur centralized in ''off-line'' events such as conferences, meet-ups, and hackathons~\cite{munir2017open, stam2009when, dahlander2005relationships}. 

Prioritization and selection of requirements are commonly performed by individuals in leadership positions of the OSS community, however, with consideration taken to expressed wishes of the community~\cite{german2003gnome, laurent2009lessons, noll2007innovation}. This hierarchy between the roles in OSS communities is often depicted with the help of an onion model~\cite{nakakoji2002evolution}. In its multi-layered construction, central and leadership roles can be found among the core layers, while the passive users can be found in the outer ones (cf. Core-Periphery Model~\cite{joblin2017classifying}). The structure implies that the further out a community member is, the less direct influence and knowledge the person has over the project's state and direction~\cite{jensen2007role}. Furthermore, what roles that exist in a community, specifically regarding leadership, may differ between communities. Some may, for example, have a project lead as with Linus Torvalds in the Linux kernel community, while some may have a core team of entrusted members as in the PostgreSQL community~\cite{nakakoji2002evolution}. 

Migration between layers can be fluid and agile depending on the project, e.g., community members can move between multiple layers, or be recruited into one, bypassing outer ones~\cite{jensen2007role}. This migration further depends on the type of governance in the community.

\subsection{Governance in OSS communities}
\label{sec:RW:governance}
\begin{figure}[htbp]
\centering
\includegraphics[width=1\textwidth]{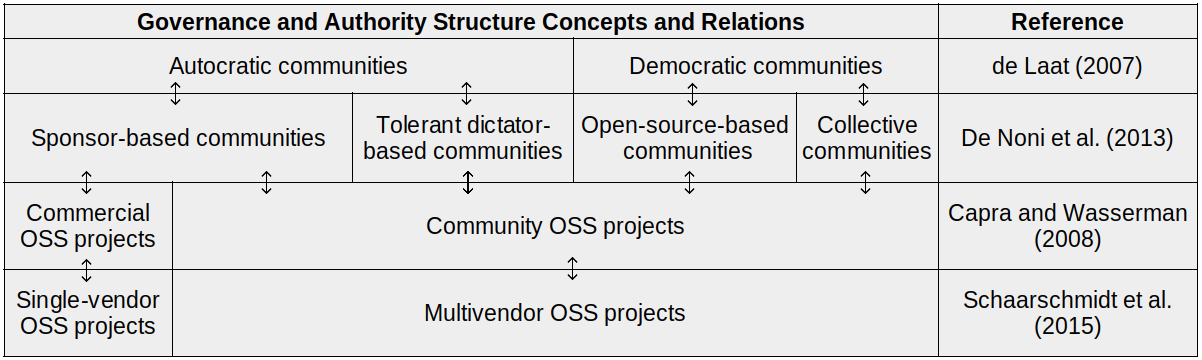}
\caption{Overview of governance and authority structure concepts in OSS projects and their relations as presented in Section~\ref{sec:RelatedWork}.
}
\label{fig:GovernanceOverview}
\end{figure}

% According to Markus~\cite{markus2007governance}, \textit{``OSS governance is the means of achieving the direction, control, and coordination of wholly or partially autonomous individuals and organizations on behalf of an OSS development project to which they jointly contribute''}. 
de Laat~\cite{de2007governance} describes OSS governance as different configurations, primarily based on the authority structure, i.e., the way that authority is established, distributed, and exercised, either through autocratic or democratic principles. In the former, leadership is centralized and top-down, while in the latter it is decentralized and bottom-up. Building on this distinction, De Noni et al.~\cite{de2013evolution} refines the two configurations further as presented in Fig~\ref{fig:GovernanceOverview}. Concerning communities with autocratic tendencies, they differentiate between sponsor-based and tolerant dictator-based communities. In the former, leadership is centered around the sponsoring firm(s), while in the latter it is centered around a single project leader (tolerant dictator). In regards to communities with democratic tendencies, De Noni et al.~\cite{de2013evolution} separates open-source-based and collective communities. In open-source based communities leadership is characterized as institutionalized, democratic, and distributed, often inside the walls of a foundation. In collective communities, leadership is seen as collective, meritocratic, and distributed. 

Capra and Wasserman~\cite{capra2008framework} makes a distinction between commercial and community OSS. In the former, the OSS project is owned and managed by a single firm~\cite{riehle2012single}, i.e., a special case of sponsor-based communities~\cite{de2013evolution}. In the latter, the community is owned and managed by the community, which may include one or more firms, also aligning with the community-managed governance model as described by O'Mahony~\cite{o2007governance}. Schaarschmidt et al.~\cite{schaarschmidt2015firms} further label these types of projects as single-vendor projects and multivendor projects respectively.

Even with the categorizations of OSS governance models and their authority structures shown in Fig~\ref{fig:GovernanceOverview}, other research shows that the picture can be more blurry. According to the literature review by Shaikh and Henfridsson~\cite{shaikh2017governing}, research has been consistent in describing how communities can only have one authority structure (with one notable exception~\cite{germonprez2014collectivism}). Even though a community can evolve its authority structure in hybrid forms with time, a single authority structure will result in the end~\cite{o2007emergence}. However, based on their view of a duality between governance and coordination, Shaikh and Henfridsson~\cite{shaikh2017governing} move to suggest that multiple forms of authority structures can co-exist in parallel, each embedded in and operationalized by a coordination process. These coordination processes can integrate, and evolve together within a community, of which some may pass out with time and be replaced by others. In their longitudinal analysis of the Linux kernel community, they identified a varying mix of autocratic and oligarchic structures, but also semi-autonomous governing in terms of the different sub-modules. Meritocracy was continuously present through the analysis. I.e., even tolerant dictator-based communities can show traits of a community-managed~\cite{o2007governance} and meritocratic~\cite{de2007governance} governance model.

Although literature lists a number of them, meritocracy may be considered one of the more common authority structures, or type of governance in OSS communities (e.g.,~\cite{butler2018investigation, scacchi2010collaboration, mahoney2005non, nakakoji2002evolution, fielding1999shared, nguyen2018do}). Based on merit and the earning of trust and status in the community, individuals are granted further responsibility and authority~\cite{fielding1999shared}. Merit correlates to the quality and quantity of the individual's contributions~\cite{german2003gnome, syeed2017measuring}. A common assumption is that these contributions are limited to technical code contributions, however, as is shown by Eckhardt et al~\cite{eckhardt2014merits}, this can be a simplification. Considering the onion model~\cite{nakakoji2002evolution}, several paths are depending on the type of role an individual possesses. Proven coordination and leadership skills are aspects that may be considered~\cite{o2007emergence, jensen2007role}, but not obviously captured in code commits. As highlighted by O'Mahony and Ferraro~\cite{o2007emergence} in their study of the Debian community, \textit{``Any examination of meritocracy must develop a context-specific understanding of how merit is conceptualized''}.

\subsection{Influencing the Requirements Engineering Process in OSS communities}

The members of the community all have their motives for participating, social or economic~\cite{riehle2010economic, lerner2002some}. It may, therefore, be considered a challenge for firms to align their internal agenda with that of the community~\cite{dahlander2008firms, schaarschmidt2015firms, omahony2008bounary}. A decision to add functionality may require consensus in the community and approval by the community leadership depending on the type of governance. Being too aggressive with one's agenda may have an adverse effect and result in the functionality being blocked~\cite{aagerfalk2008outsourcing}.

Dahlander et al.~\cite{dahlander2005relationships} differentiate how firms can adapt their relationship with an OSS community based on the level of influence needed. On a continuum scale, a relationship can be characterized as parasitic, commensalistic or symbiotic. In the parasitic approach, the firm takes without giving back, by some referred to as a ``free-rider''. In the commensalistic approach, the firm contributes back when motivated, but focus on internal development. In the Symbiotic approach, the firm also sees to the best of the community, working to align internal and external development. The alignment is created through working as peers, and building status and recognition inside the community~\cite{dahlander2006man}. 

To build a symbiotic relationship, firms should first understand and learn to respect the needs, norms, and structure of the community~\cite{butler2018investigation, nguyen2017coopetition, dahlander2005relationships, dahlander2006man, aagerfalk2008outsourcing, lundell2010open}, a form of ``good citizenship''~\cite{omahony2008bounary}. If there is a foundation encapsulating the OSS community, firms may have the option to gain influence through membership or sponsorship~\cite{mahoney2005non, omahony2008bounary}, or in other ways supporting the foundation, e.g., by supporting development with infrastructure~\cite{dahlander2005relationships}, or general subject matter expertise~\cite{butler2018investigation}. In return, they may receive seats at relevant boards and committees through which they can make their voice heard~\cite{butler2018investigation, mahoney2005non}. Foundations, and similar boundary organizations between firms and an OSS community, are often limited to managing the technical direction of an OSS projects~\cite{omahony2008bounary}.

A more direct and general approach to the control of code contributions is by having ``a man on the inside'', letting employees engage with the community~\cite{henkel2008champions, dahlander2006man, schaarschmidt2015firms, nguyen2017coopetition, nguyen2018do, omahony2008bounary}. An alternative is to contract members of the community directly to have them work on matters of importance to the firm~\cite{dahlander2008firms, german2003gnome, schaarschmidt2015firms, riehle2011controlling, omahony2008bounary}. Through their engagement, these sponsored community members can take part in the RE processes by participating in discussions and providing both technical and non-technical contributions and support~\cite{munir2017open, butler2018investigation}. This work may take place both online and offline, because being visible and active on both ends is essential ~\cite{stam2009when, o2007emergence, munir2017open, schaarschmidt2015firms}. 

% According to Schaarschmidt et al.~\cite{schaarschmidt2015firms}, this approach to gaining influence through active engagement can be divided into two categories, control by leadership, and resource deployment control. In the former, influence is obtained by having employees in leadership positions of a community. In the later, influence is gained by having employees work in the community and infusing the community with the firm's norms and values.

\subsection{Determining the need for Influence in OSS communities}

As highlighted by Dahlander and Magnusson~\cite{dahlander2008firms}, it may be difficult to determine which OSS communities are of strategic importance to their operations. Firms should identify how they could benefit from an OSS project and its community, and then what kind of engagement is required to reap these potential benefits~\cite{butler2018investigation}. 

From a business model perspective, it may be considered how the OSS project helps to create, deliver, and capture value for a firm~\cite{teece2010business}. It may, for example, serve as a basis on which the firm builds complementary products or services, such as support and subscription offerings, or proprietary extensions~\cite{sharivar2018business}. The OSS project could also function as a product or service enabler, embedded in hardware products~\cite{linaaker2018motivating}, or as tooling and infrastructure for development and service delivery~\cite{munir2017open}. From a more strategic perspective, the OSS project may provide value as a foundation for pooled R\&D/product development, and as a mean for standardization of technology~\cite{west2006challenges}. Furthermore, just as the community may serve as an external workforce, it may also serve as a marketing channel, both for customers and future employees~\cite{henkel2006selective, dahlander2008firms, riehle2011controlling}. Hence, the value should be viewed both from a monetary and a non-monetary perspective~\cite{sharivar2018business}.

From a technical perspective, it is also essential to understand the strategic connection of the OSS project to a firm's business and how this is reflected in a developer's level~\cite{butler2018investigation}. There may be internal dependencies and integrations between the OSS project and internal software that are critical to maintain~\cite{munir2017open}, as is specific functionality that is requested and expected by the firm's customers~\cite{linaaker2018motivating}. These two reasons both warrant a need for alignment between software development inside the firm and the community respectively~\cite{dahlander2008firms, schaarschmidt2015firms}. If the direction of the community is predictable, both regarding road-map and release planning, then the need for an active community presence may be less urgent~\cite{butler2018investigation}. 

\section{Research Design}
\label{sec:ResearchDesign}
%==============================================
%==============================================

\begin{figure}[htbp]
\centering
\includegraphics[width=1\textwidth]{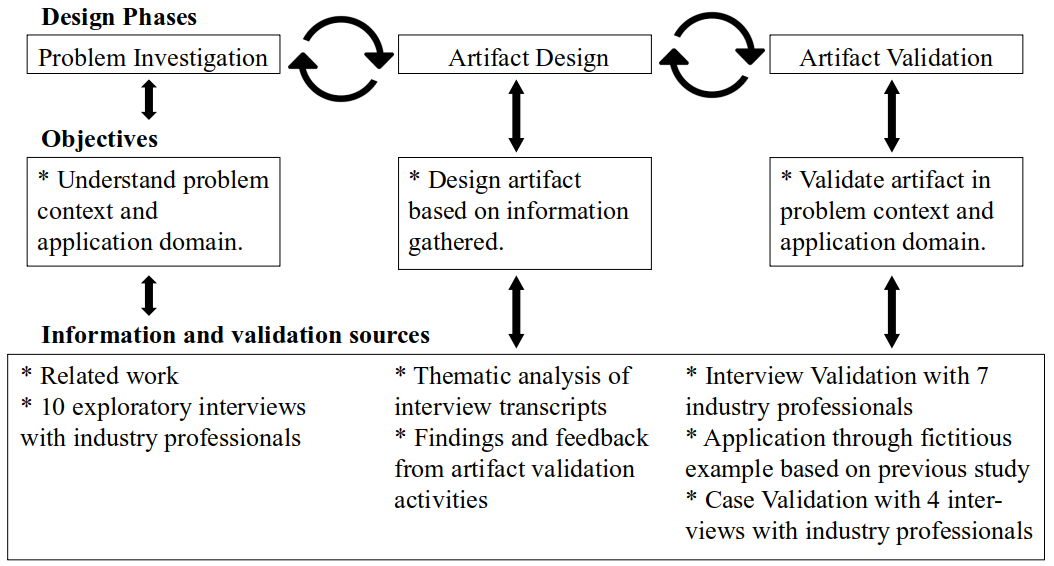}
\caption{Overview of the research process and context used in this study, using design science research~\cite{wieringa2014design, hevner2004design}. Design steps includes problem investigation, artifact design and artifact validation, which are performed iteratively.}
\label{fig:ResearchProcess}
\end{figure}

To develop the CSF, we used a design science research approach~\cite{wieringa2014design, hevner2004design}, in which research is performed and structured in the form of design cycles. A design cycle is comprised of three phases: problem investigation, artifact design, and artifact validation~\cite{wieringa2014design}. These phases are performed iteratively, as exemplified in Figure~\ref{fig:ResearchProcess}. For example, as artifact validation renders feedback, this feedback is used for refinements in artifact design, resulting in a new artifact design that needs validation. Below, we use this structure to describe how we planned and executed the research behind this study.

% The research is based on design science~\cite{hevner2004design, wieringa2014design}, and continues from previous work which constituted a first design cycle~\cite{linaaker2017contribution}. In this previous design cycle, a first version of the CSF was designed based on a literature survey and an expert interview. In this study, we present a second design cycle where we leverage a series of ten semi-structured expert interviews with a questionnaire based on the first version of the CSF. Interview transcripts were coded with a thematic approach which resulted in a new version of the CSF. This new version was then validated through further seven further expert interviews where the interviewees were presented with the CSF and asked questions regarding its completeness and correctness. To further validate against these quality criteria, in one of these interviews, the framework was also applied to a fictitious example of an earlier reported case study~\cite{munir2017open}, which is presented in this study.

% Literature review and expert interview provided theoretical underpinning... Remove Earlier work as a framework

% continue from an earlier design cycle which results have been published in earlier work~\cite{linaaker2017contribution}. Based on reviewed literature and an expert interview, a first version of the CSF was presented. With this background, and to analyze the problem context further,

\subsection{Problem Investigation Phase}
In the problem investigation phase, the problem context is analyzed. In our case, we conducted exploratory interviews to understand industry practice beyond what has been identified in the literature. 

Ten individuals were interviewed (denoted I1-10, see Table~\ref{tbl:Interviewees}) with a semi-structured approach where the interview instrument consisted of open-ended questions (see Section 9, Appendix A). The interviewees all held positions with responsibilities relevant to understanding how their respective firms work and engage with OSS communities. They were selected based on convenience sampling. All interviews lasted between 30 to 60 minutes and were conducted either in person or over video link by the first author of this study. All interviews were audio recorded and transcribed.

% with the purpose of helping \textit{`` \ldots firms engaged in Open Source Software (OSS) ecosystems to create contribution strategies which motivate what they should contribute and when, but also what they should focus their resources on and to what extent''}~\cite{linaaker2017contribution}. The former part, referring to what should be contributed and when is further addressed in a separate study which focuses on the construction of contribution strategies for software artifacts~\cite{linaaker2018motivating}. The latter part is what this study aims to address, which is further articulated in \textbf{RQ1} and \textbf{RQ2}. 

% To analyze the problem context further, a series of exploratory interviews were conducted in order to understand industry practice beyond what has been identified in the literature. Ten individuals were interviewed (denoted I1-10, see Table~\ref{tbl:Interviewees}) with a semi-structured approach where the interview instrument consisted of open-ended questions based on the structure of the initial framework which has been reported earlier~\cite{linaaker2017contribution}, and the extended literature review presented in Section~\ref{sec:RelatedWork}. The interviewees all held positions with responsibilities concerning how their respective firms work and engage with OSS communities. All interviews were audio recorded and transcribed.

\subsection{Artifact Design Phase}
Drawing on the knowledge and understanding that is obtained during the problem investigation, an artifact is designed with the hypothesis that it will address the design problem. In this study, the design problem is stipulated by \textbf{RQ1} and \textbf{RQ2}, and the artifact is the CSF. 

% The purpose of the framework is to help firms create and tailor community strategies that enable them to focus resources on communities that matter and gain the influence needed on their respective RE processes.

% The interviews in the problem investigation phase of this study were semi-structured, and all followed the same questionnaire (see Section 9, Appendix A) which allowed for the researcher to keep interviews to the same line of topics, while still giving room to explore when needed. 
% Also, all interviewees were given the same background explanation concerning the research topic and how we define central concepts such as influence and requirements engineering.
% Audit trails~\cite{runeson2012casestudy} were kept to enable a systematic mapping of the data generated from interviews. 
% This mapping includes the interview transcripts and the different levels of abstractions and groupings that were performed during the thematic analysis process~\cite{cruzes2014case, cruzes2011research}. 
% Furthermore, in the validation phase of the design process used in this study, member checking~\cite{easterbrook2008selecting} was performed as interviewees from the problem investigation phase were used to validate if the researchers understanding, captured in the CSF, was valid. Four new interviewees were also included to bring an outside, unbiased perspective on the framework and its content. 

The interview transcripts were coded with an inductive approach by the first author with audit trails intact~\cite{runeson2012casestudy}. Sentences and paragraphs were first assigned descriptive topics. These topics were later collected under common codes, which could then be related and sorted under \textbf{RQ1} and \textbf{RQ2} respectively. Codes relating to \textbf{RQ1} are referred to as aspects and are divided into three categories; Business Aspects (BA), Technical Aspects (TA) and Community Aspects (CA). Codes relating to \textbf{RQ2} are referred to as engagement practices and are collected in one single category. The CSF is presented in full detail in Section~\ref{sec:CommunityStrategyFramework}

Below we provide an example with a subset of quotes rendering in engagement practice 5 (EP5) of the CSF: 
\begin{itemize}
    \item Engagement Practice (RQ2)
        \begin{itemize}
        \item Offer the expertize and resources of the firm
            \begin{itemize}
            \item Quote by I1: \textit{`` \ldots contributing DevOps-kind of information and documentation and information, and it gives credebility''}.
            \item Quote by I7: \textit{``We don't have developers, but we send you these machines to do testing''}.
            \end{itemize}
      \end{itemize}  
\end{itemize}

\begin{table*}%[htbp] %BR
\footnotesize%BR
\centering
% \caption{Ten industry professionals (I1-I10 from nine different firms were interviewed in the problem investigation phase. Seven industry professionals (I1, I5, I6, I11-I14) were interviewed in the validation phase. Small-sized firms (S): \textless 50 employees, Medium-sized firms (M): 50 \textless \textgreater 250 employees, Large-sized firms (L): \textgreater 251 employees)}
\caption{Ten industry professionals (I1-I10) were interviewed in the problem investigation phase. Seven industry professionals (I1, I5, I6, I11-I14) were interviewed in the Interview Validation phase. Four industry professionals (I15-18) were interviewed in the Case Validation. Small-sized firms (S): \textless 50 employees, Medium-sized firms (M): 50 \textless \textgreater 250 employees, Large-sized firms (L): \textgreater 251 employees)}
\label{tbl:Interviewees}
%\begin{tabular}{p{0.2cm} p{3.3cm} p{0.6cm} p{1.7cm} p{0.6cm} p{3.3cm}}
\vspace{0.5em}\begin{tabular}{p{0.2cm} p{3.2cm} p{0.6cm} p{1.6cm} p{0.4cm} l}
\toprule
% ID & Title & Firm & Description & Size & Use of OSS \\ \midrule
% I1 & OSS Program Officer & A & Telecom & L & OSS in Infrastructure \\
% I2 & Community Manager & A & Telecom & L & OSS in Infrastructure \\
% I3 & OSS Program Officer & B & Software products & L & OSS in Infrastructure and products \\
% I4 & OSS Strategist & C & Software products & L & OSS in Infrastructure and products \\
% I5 & Community Manager & D & Software products & S & OSS products \\
% I6 & Community Manager & E & Software products & S & OSS products \\
% I7 & OSS Strategist & F & Software products & L & OSS products \\
% I8 & OSS Program Officer & G & Software products & L & OSS in Infrastructure and products \\
% I9 & OSS Program Officer & H & Software products & L & OSS in Infrastructure and products \\
% I10 & OSS Strategist & I & Consumer electronics & L & OSS in products \\
% I11 & OSS Strategist & J & Consultancy & S & OSS strategy services \\ 
% I12 & Community Manager & F & Software products
%  & L & OSS products \\ 
% I13 & Community Manager & F & Software products
%  & L & OSS products \\ 
% I14 & OSS Program Officer & K & Consumer electronics & L & OSS in products \\ \bottomrule

ID & Title & Firm & Business & Size & Use of OSS \\ \midrule
I1 & OSS Program Officer & A & Telecom & L & Infrastructure  \\~\\
I2 & Community Manager & A & Telecom & L & Infrastructure \\~\\
I3 & OSS Program Officer & B & Software products & L & Infrastructure \& Products \\
I4 & OSS Strategist & C & Software products & L & Infrastructure \& Products \\
I5 & Community Manager & D & Software products & S & Products \\
I6 & Community Manager & E & Software products & S & Products \\
I7 & OSS Strategist & F & Software products & L & Products \\
I8 & OSS Program Officer & G & Software products & L & Infrastructure \& Products \\
I9 & OSS Program Officer & H & Software products & L & Infrastructure \& Products \\
I10 & OSS Strategist & I & Consumer electronics & L & Products \\
I11 & OSS Strategist & J & Consultancy & S & Strategy services \\~\\ 
I12 & Community Manager & F & Software products
 & L & OSS products \\ 
I13 & Community Manager & F & Software products
 & L & Products \\ 
I14 & OSS Program Officer & K & Consumer electronics & L & Products \\ 
I15 & Team Manager & L & Embedded systems & L & Infrastructure \\  
I16 & Project Manager & L & Embedded systems & L & Infrastructure \\  
I17 & Senior Developer & L & Embedded systems & L & Infrastructure \\  
I18 & Junior Developer & L & Embedded systems & L & Infrastructure \\ 

\bottomrule
\end{tabular}
\end{table*}

\subsection{Artifact Validation Phase}
In the artifact validation phase, the artifact is tested as a candidate solution to the defined design problem. In our study, this phase consisted of three steps. First, we conducted seven validation-focused interviews with four new industry professionals (I11-14), but also three from the problem investigation phase (I1, I5, I6), see Table~\ref{tbl:Interviewees}. Second, to evaluate applicability and utility (i.e., descriptive validation~\cite{hevner2004design}), the framework was applied through a fictitious example on a previously performed case study on how Sony Mobile evolved in their engagement with the communities of Jenkins and Gerrit~\cite{munir2017open}. Third, the CSF was validated in a similar way as in the first step, but within the context of a software-intensive firm (CaseOrg) and its Tools department.

\subsubsection{Interview Validation}
\label{subsubsec:InterviewValidation}
The CSF was presented and discussed one element (aspect or practice) at a time to the interviewees. Discussions focused on whether something was redundant, missing, or could potentially be modified. Interviews were audio recorded and transcribed. 

After verifying with transcripts, this step of the validation phase resulted in one TA being removed, TA2 being reformulated to also focus on the ''fitness-of-use'', and the explicit addition of TA4. A further consideration brought up in several of the validation interviews was that, while the business and technical aspects are relevant for determining the need for influence on the RE process, the community aspects are on the other hand used for determining the feasibility and potential of gaining influence in the community. I13, for example, describes it as, \textit{``So your first two sets of criteria, the business and technical aspects, felt like you were deciding yes or no, we should care about this community. The community aspects don't feel like yes/no's, we should care, these feel much more like feasibility, can we do it or not''}. Furthermore, the validation interviews resulted in the validation of and more nuances to existing aspects and practices. I12 for example added to BA2 the perspective that standardization can be part of a strategy to build a software ecosystem. From a design science perspective, findings and feedback from the validation phase were used to refine the artifact design.

\subsubsection{Framework application example}
\label{subsubsec:frameworkApplciationExample}
The analysis was performed by the first author of this study, who was also one of the authors behind the previous case study~\cite{munir2017open}. Using interview transcripts and codings from the original study, the CSF was applied by considering each aspect against the Jenkins and Gerrit communities. Traces and support for the different engagement practices were then searched for. Findings were then summarized and verified for correctness with the OSS Program manager at Sony Mobile. The program manager was presented with the results from each of the applied practices, and the support gathered for each of the engagement practices. The program manager was asked to verify the interpretation and clarify any misunderstandings of the first author's analysis. It should be noted that the program manager was also one of the interviewees from the previous case study~\cite{munir2017open}. 

\subsubsection{Case validation}
\label{subsubsec:caseValidation}
The Tools department has a similar organization and purpose as that described in earlier work of Sony Mobile, which is the foundation for the application example as described in Section~\ref{subsubsec:frameworkApplciationExample}. CaseOrgs's Tools department develops and maintains multiple OSS tools and infrastructure projects, including Jenkins and Gerrit, to support its product development organization. All OSS communities that were discussed during interviews were characterized as community-managed and meritocratic. 

Four interviews were conducted with I15-I18 (see Table~\ref{tbl:Interviewees}) who all held various positions but were all engaged in different OSS communities, some with maintainership positions. As in the previous step (see Section~\ref{subsubsec:InterviewValidation}), the CSF was presented and discussed one element (aspect or practice) at a time to the interviewees. Discussions focused on whether something was redundant, missing, or could potentially be modified, specifically in the context of the communities that CaseOrg's Tools department is engaged in. Interviews were audio recorded and transcribed. Findings and feedback were used to refine the artifact design of CSF. No aspects or practices were removed or added. Existing ones were however given more nuances as EP5 where the importance of attending and arranging hackathons was added.

% For a further initial validation of the CSF, we applied through a fictitious example on a previously performed case study~\cite{munir2017open}. The case concerns Sony Mobile and their engagement in the Jenkins~\footnote{https://jenkins.io/} and Gerrit~\footnote{https://gerritcodereview.com/} OSS communities and is based on earlier research where we have analyzed this engagement in detail. The analysis from the example was presented to and discussed with the Open Source Programs officer at Sony Mobile in order to verify consistency and correctness in how the case was interpreted.

%==============================================
%==============================================
\section{Community Strategy Framework}
\label{sec:CommunityStrategyFramework}
%==============================================
%==============================================

\begin{figure}[b!]%[t!]
\centering
\includegraphics[width=1\textwidth]{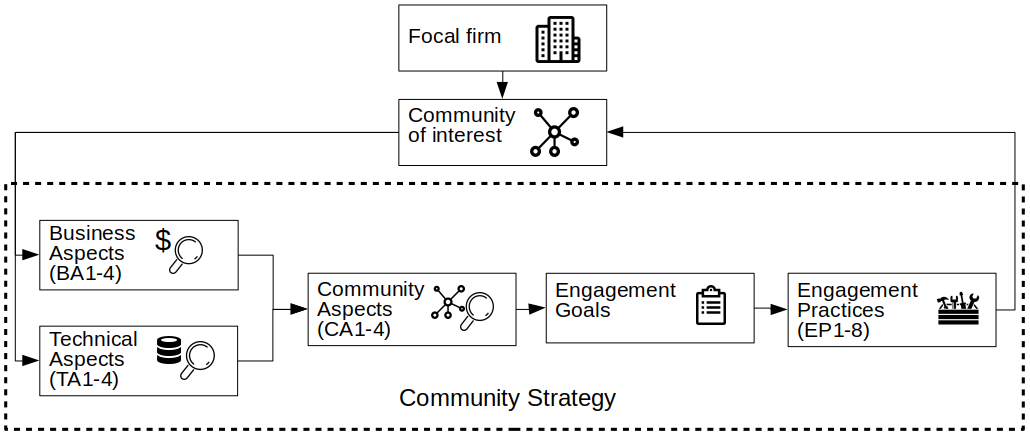}
\caption{Overview of the Community Strategy Framework's related process. A firm first values the community of interest with the business and technical aspects and then uses the community aspects to determine the feasibility of gaining influence and potential engagement goals. Engagement goals are then decided and engagement practices chosen.}
\label{fig:frameworkProcess}
\end{figure}

Here we describe the Community Strategy Framework (CSF) as presented in Table~\ref{tbl:CommunityStrategyFramework}, which consists of two parts. The first of these, contain aspects a firm should consider when assessing its need to influence the RE process in an OSS community (\textbf{RQ1}). The second part of the CSF consists of practices a firm should consider to gain influence on the RE process in an OSS community with meritocratic governance or aspects thereof~\cite{shaikh2017governing} (\textbf{RQ2}). Figure~\ref{fig:frameworkProcess} shows an overview of the CSF. A firm constructs a community strategy by firstly assessing the community of interest based on four Business Aspects (BA1-4) and four Technical Aspects (TA1-4), and secondly determine the actual need for and feasibility of gaining influence using the four Community Aspects (CA1-4). It may be that not all aspects are applicable or relevant. It may also be that one aspect may indicate a need for influence, while another may not. With this in mind, it is up to the user to consider the different aspects in relation to the community of interest, and weigh these against each other. The CSF should, therefore, be viewed as a support for the user to arrive at a decision on if and how much influence is needed by the firm on the RE process in the OSS community.

Once such a decision has been made, the firm then formulates important engagement goals and selects which Engagement Practices (EP1-8) to apply, and finally determine how to apply them. Below we present the respective aspects and engagement practices in detail.

For further guidance on how to apply the CSF, please see Section~\ref{sec:CaseExample} where it is applied in a case example based on earlier work~\cite{munir2017open}.

\begin{table}[t!]
\footnotesize
\centering
\caption{Overview of the Community Strategy Framework. Business, Technical and Community Aspects relate to \textbf{RQ1} and Engagement Practices to \textbf{RQ2}}
\label{tbl:CommunityStrategyFramework}
\vspace{0.5em}\begin{tabular}{p{0.7cm} p{10.8cm}}
\toprule
\multicolumn{2}{l}{\textbf{Business Aspects (BA)}} \\ \midrule
BA1 & Connection between the OSS project and the value proposition and revenue streams of the firm's business model \\
BA2 & Connection between the OSS project and the business strategy of the firm \\
BA3 & Importance of the OSS community as a pool for recruitment \\
BA4 & Need of the OSS community-related visibility and credibility towards the firm's customers \\ \midrule
\multicolumn{2}{l}{\textbf{Technical Aspects (TA)}} \\ \midrule
TA1 & Internal dependency of the OSS project inside the firm \\
TA2 & Fitness-of-use and road-map alignment of the OSS project \\
TA3 & Dependency on the OSS community's release planning \\ 
TA4 & Need for competence and resources of the OSS community \\ \midrule
\multicolumn{2}{l}{\textbf{Community Aspects (CA)}} \\ \midrule
CA1 & Presence, influence and agenda of other stakeholders in the OSS community \\
CA2 & Diversity and activity in the OSS community's stakeholder population \\
CA3 & Openness in Culture and Governance of the OSS community \\
CA4 & Ownership and management of the OSS project \\ \midrule
\multicolumn{2}{l}{\textbf{Engagement Practices (EP)}} \\ \midrule
EP1 & Understand the governance structure and have seats in right groups, committees and boards \\
EP2 & Become a member or sponsor of the foundation or governing community body \\ 
EP3 & Sponsor, contract or hire developers and maintainers to engineer contributions and mentor internal engineers \\
EP4 & Contribute to the development of the OSS project through internal engineers\\
EP5 & Offer the expertize and resources of the firm \\ 
EP6 & Have an active on-line and off-line community presence \\
EP7 & Be open and humble to the OSS community \\
EP8 & Build an inner source culture and practice inside the firm \\ \bottomrule
\end{tabular}
\end{table}

\subsection{Aspects}

The aspects are divided into three categories: business, technical, and community aspects. Aspects from the two former categories are used to reflect on the OSS and its importance to the firm from a business and technical perspective. 
%I11 gives his interpretation as \textit{`` ''these [business aspects] are the open source boxes that we're using in our business, these five boxes'' and the executives don't know what's in these boxes, and the technical aspects are looking inside these boxes''}.
The latter, community aspects, are used to reflect on the feasibility and potential to gain influence, as well as the need for it.

% JB: I think these are the right dimensions that you're looking at it because... There is an assessment, in my mind, it's kind of two layers of this right? You got the business value assessment of open source, which I think you've covered in the business aspects very well. And then there is... "These are the open source boxes that we're using in our business, these five boxes" and the executives don't know what's in these boxes, and it sounds like these technical aspects are looking inside these boxes.

% JB: Of these three aspects (categories), I would argue that the community aspects are a lot less important than the business and technical aspects. Because I think most frankly most companies aren't interested in the community aspects, they may be to some degree, but I think it's a much smaller set of companies. But every company has to consider the business and technical aspects of open source.

% SP: So you're first two sets of criteria, the business and technical aspects, felt like you were deciding yes or no, we should care about this community. This set doesn't feel like yes/no's, we should care, this one feels much more like feasibility, can we do it or not. And I imagine some community aspects are like yes/no we should be involved, but these are all like - after you've decided, can you actually make an impact.

\subsubsection{Business Aspects (BAs)}

\textbf{BA1 - Connection between the OSS project and the value proposition and revenue streams of the firm's business model}.
As expressed by I10, \textit{``It comes down to the bottom-line, and making sure where [the firm] is making money, we want to have as much impact on those areas as possible''}. I11 emphasizes \textit{``I think the sticky point is understanding how the value of the open source matches to the value of the business they're trying to build''}. A firm should, therefore, recognize how the OSS project is leveraged in its business model. It can be a complement of the core value proposition as for Red Hat and their distribution Red Hat Enterprise Linux which is based on Fedora. It could also be an enabler for the value proposition as for Sony Mobile and Android which is used in their mobile handsets. Or it could play a more indirect role as part of an infrastructure or a tool-chain that can be used to develop and deliver the main value proposition. In the latter case, there may be a limited amount of competitive edge connected to how the OSS project is used internally, as described by I16, \textit{``We're so far out of core business that we have our own contribution process''}. In other cases, \textit{``if you're building a product offering around an open source project in the core, there's not even a question. If you're committing to customers to support the project, then you need to have an influence on that project''} (I12).

\textbf{BA2 - Connection between the OSS project and the business strategy of the firm}.
The business strategy specifies how a firm should navigate a changing environment and as a consequence construct and adapt its business model~\cite{dasilva2014business}. In this context, an OSS project and its community can play a pivotal part, e.g., to commoditize a market, or change the default technology being used by industry. I9 reflected on one of their experiences, \textit{``So we wanted to change the industry conversation, and we wanted to have a substantial impact in that''}. This type of standardization can further be part of a strategy where the intent is to build a software ecosystem, as explained by I12, \textit{``Driving standardization enables the market to potentially develop and that is what gives business opportunity, if you're running an infrastructure project and all of a sudden you have a lot of third-party vendors, whether monitoring and logging, or storage or network, you know there's an entire ecosystem that comes along, not to mention all of the developer toolings that you need to develop container-native applications. So there's a lot of opportunities that come from having a de facto technology base in the platform. And that creates that opportunity to create the commercial ecosystem around the platform''}.

% Standardization allows firms to shift focus to work on differentiating technology.
% Change BA2 to standardization/commoditization?
% Potential for market disruption? Bringing down substitutes 
% Potential to create complements? When the price of a good goes down, demand for complements will go up
% Part of a upcoming or current standard?

\textbf{BA3 - Importance of the OSS community as a pool for recruitment}.
Being active and influential in an OSS community can be important to attract and maintain a skilled workforce. This concerns both specific technologies where skilled people are scarce and attracting developers in general. The latter is emphasized by I7, \textit{``It's also about reputation - [firm] created a big open source office, and a huge open source initiative, because no one wanted to work with them''}. I11 adds, \textit{``This is something a lot more are starting to realize, particularly large companies with aging populations, that people don't want to sit in a stodgy old company in cubicles''}. I16 continues, \textit{``We need to show that we don't just consume, but also contribute to attracting good developers. The community becomes a channel for new employees''}.

\textbf{BA4 - Need of the OSS community-related visibility and credibility towards the firm's customers}.
As for recruiting talent, being active and influential in an OSS community can be essential to attract and maintain customers. It can be to prove technical competence, but also the ability to push features upstream. As put by I1, \textit{``[The OSS project] was the selling point of the product. We needed to demonstrate to customers that we were one of the core contributors of [the OSS project]''}. I11 gives the example of IBM and how they, \textit{`` \ldots back in the 2000s, invested a billion dollars in Linux and they wanted to make a big deal of it because they saw that as an emerging market that they wanted to get into''}. I1 adds how this aspect is particularly important for firms using OSS in their products, such as \textit{``Red Hat, or any commercial open source project. Like Cloudera would need to do that, that they have influence in the Hadoop community, and DataStack for Cassandra''}.

\subsubsection{Technical Aspects (TAs)}

% \textit{TA1 - Importance of the OSS Project to the firm's developers}
% An OSS project can be important for the firm's developers as explained by I1, \textit{``It keeps my developers happy to work on this technology, and to refine their skill-set because people are not necessarily willing to stay at an employer where they are working on old tech constantly. They will get bored, and worst case, recognize the market opportunity for someone who is excellent at this technology''}. Therefore, if certain developers are of specific value, the firm should be careful about choosing alternative OSS projects. This is further emphasized by I12, \textit{``You can't take a maintainer on an OpenStack module and put him on working on something else. Because there's an allegiance to a mission, a set of values, and also the technology in the code, right? The developers feel ownership of the code they write, that's especially true in the open source world''}.

\textbf{TA1 - Internal dependency of the OSS project inside the firm}.
Technical dependencies between an OSS project and a firm's internal software can be considered an architectural reflection of how an OSS project connects to a firm's value proposition (BA1). Certain features in a product or parts in an infrastructure may be dependent on the project. I3 phrases the question as \textit{``Do you depend on this or do you not? And how much do you depend on it? How much functionality goes into your product that's based on upstream software? How heavily are these integrated into your product?''}. I17 exemplifies, \textit{``We are extremely dependent on [OSS project], we have based our whole infrastructure chain on it. This requires us to be active so that we can affect in what directions the tools head''}.

\textbf{TA2 - Fitness-of-use and road-map alignment of the OSS project}.
Deviance between a firm's internal and a community's requirements and road-maps may be essential to address in order to avoid or minimize technical debt. I12 refers to an OSS project's fitness-of-use and explains it as \textit{``How many things that we need it to do does this project do today? And if you feel like there is a delta between what it does today, and what you need it to do, and this is a strategically important component of your plan, then it would be important to be involved. You need to have influence so that you can affect the change that you need in that project''}. I1 adds that in \textit{`` \ldots some cases, it may not be necessary for you to be as actively involved because you are happy with the direction it is going, and it's a mature and stable project. And in some cases, you really need to be there and watch it and make sure it goes in the right direction''}.

\textbf{TA3 - Dependency on the OSS community's release planning}.
A firm can be more or less dependent on the release planning of an OSS community, and have various needs to synchronize it with that of any internal development. Getting features upstream quickly and running the latest release may be an essential factor for firms whose customers may expect quick access to the latest functionality, as some buyers do of Android-based mobile handset manufacturers. For others, it may be less of a concern, as for Red Hat who focuses on offering a stable and secure version of Fedora. I4 explains it as \textit{``How much do we care if they are changing it rapidly? Are we living on a fork and are willing to eat a little bit of fit and finish? Or do we really want to be on the latest bits all the time?''}. I12 sees it from a risk analysis perspective, \textit{``Is there is a risk that a feature will not go into a project, or is there a risk that a project that you depend on will miss its release date?''}.

\textbf{TA4 - Need for competence and resources of the OSS community}.
Firms can be limited, both in terms of \textit{`` \ldots resources, time or people''} as highlighted by I16, or in terms of specific competencies that are internally available. By engaging in an OSS community, firms may have an opportunity to gain these resources through collaboration and ``co-opetition''. By growing influence in such a community, a firm can better exploit and steer these resources to best match the firm's agenda. I1 explains it as, \textit{``Sometimes we may not have the competency inside the company, but yet we want to draw on the competency of the community to help us use something correctly. So, the link to the community may be important because of that, to just improve our own competency in handling that project''}.

\subsubsection{Community Aspects (CAs)}

\textbf{CA1 - Presence, influence, and agenda of other stakeholders in the OSS community}.
Knowing whom the stakeholders are, where they focus their resources, with whom they collaborate and how much influence they hold, can signal how a firm should consider its relationship with the stakeholder, but also overall community engagement. As expressed by I3, \textit{``If it's a company that wields a big influence, and they are a competitor to you, there's a much different way to approach that than if they were a partner or one you're not a competitor with''}. I3 continues, \textit{``If you understand why those companies are contributing it potentially makes your strategy why you should be contributing''}. I1 explains that for projects which are important to a firm, \textit{``You work on it, you contribute to it, and you make sure your competitors are not influencing it differently than you would''}. Hence, the presence of competitors may indicate \textit{``how strongly [a firm] need to be present''}, as further highlighted by I1. However, as explained by I13, OSS communities provide a \textit{`` \ldots forum for competitors to cooperate in a way that doesn't upset their shareholders, a form of co-opetition''}. Presence of competitors may also be \textit{``a signal that the OSS project we should be engaged in, maybe it is becoming an industry standard''}, as suggested by I2.

\textbf{CA2 - Diversity and activity in the OSS community's stakeholder population}.
A community maintained by only a few individuals or companies could be vulnerable if they were to leave for any reason. As asked by I3, \textit{``What would be the case if there were shortages in supply, i.e., the project would no longer be available?''}. I12 compares a community to an external vendor and asks, \textit{``Is this company going to be in business in five years? Is there someone who can take up the mantle if they shut down?''}. A low level of diversity and activity in a community can, therefore, be a warning sign if a firm is to engage in the first place. However, if a firm is dependent on a community or sees potential, then it indicates that the firm should invest, \textit{`` \ldots not for influence, but for health''} as emphasized by I1. I9 further adds, \textit{``We want to make sure it is not just totally dependent on one or two parties only because vibrant community to me means that it has a broad spectrum of contributions and that it is not just totally dependent on one party''}. From a sourcing perspective, it may be relevant to also consider other alternatives and weigh these against the cost of investing in the concerned community.

\textbf{CA3 - Openness in Culture and Governance of the OSS community}.
To be attractive, the culture and governance of an OSS community should have meritocratic influences, i.e., be open to new members joining and gaining in rank, but also for discussions regarding road-maps and ways of working to be open. This is further explained by I9, \textit{`If there are communities that are uninterested in changing and learning then that is a community in my opinion that will stagnate and contract, they will have a hard time growing new leadership, they will have a hard time evolving as the needs of users evolve, as the needs of community evolve''}. I9 continues, \textit{`An openness to constant improvement and to input needs to be a core value, or at least be demonstrated in a community in order to consider putting any substantial investment''}. If the project is important for the firm, a low level of ''openness'' could motivate a high investment and active engagement to be able to affect the culture and governance of the community if deemed possible.

\textbf{CA4 - Ownership and management of the OSS project}.
A criterion before engaging in an OSS community is to determine whether there is potential for the firm to gain influence and extract the expected value. As highlighted by I10, \textit{``If we see that it's a project that is controlled by one company, and it doesn't look like we'll be able to influence it in a way we want, we may not get involved in that project''}. I.e., if the OSS project is now owned and managed by the community, or a legal entity representing it (e.g., a foundation), gaining influence through active contributions and engagement may prove hard. If a firm's strategy is to hire a maintainer to get influence and there is no one available, \textit{`` \ldots the project becomes much less attractive''}, as stated by I10.

\subsection{Engagement Practices (EPs)}

The engagement practices presented in this section should be seen as a tool-box of ways in how a firm can engage with a meritocratic community to build the influence needed.

\textbf{EP1 - Understand the governance structure and have seats in right groups, committees, and boards}.
Depending on the complexity of the community governance structure, there can be many groups and committees where decisions are made. As described by I13, \textit{``It's very dependent on the community, some have large foundations, while others may have less''}. Hence, a firm should first \textit{``\ldots understand where decisions get made and what kinds of decisions [they] need to influence. Is it a technical decision? Is it a positioning decision? Is it a communication decision? And hence, which body do you need to be on, or what level of membership do you need?''} as stated by I1. 
Once understanding the governance structure of the community, a firm may need to build a certain level of influence to be able to join the identified groups. This need can also concern groups and committees that may not be a direct part of the community, but part of a greater ecosystem affecting the OSS community. As expressed by I1, \textit{``It's an influence game making sure you have people in all the right places, joined all the right foundations''}. I2 provides an example, \textit{``I think with [OSS community] they did that, ok, we need to sit at this group, this group, this group, we need to get a seat at the table of the user committee, we need to be on this committee, and they actually mapped it, and they put people there''}.

\textbf{EP2 - Become a member or sponsor of the foundation or governing community body}.
Once a firm understands the community and its governance structure, they can start to consider whether they should become a member or sponsor if possible. If the community is run under a foundation, a membership can give a firm visibility and marketing to show community, customers, and potential newly-hires that they are both competent and committed in regards to an OSS community. However, it does not have to imply a direct influence on the OSS community automatically. As explained by I3, \textit{``You can potentially buy yourself into the business side of the governance, but you don't get any technical influence unless you do any work''}. I13 gives the example of GNOME, \textit{``You pay to be part of the advisory board, but it has very little power. You have to be a contributing member to be elected to the board of directors, and that's where the power is''}. A membership or sponsorship should instead be seen as a long-term investment that can help build a sustainable influence through growing and attracting influential community members and maintainers. Sometimes membership may be unnecessary, as explained by I1, \textit{``A lot of these bodies have end-user boards which do not require any pay-to-play, it just requires you to be a big user of that technology. Because a lot of projects are very eager to get feedback on how you are using it, what are the challenges that you face at scale? So they see that as currency and value. So we've kind of been reexamining our presence in some of these bodies and asking why are we spending 40K when we can get the same influence through being on the end-user committee?''}.

\textbf{EP3 - Sponsor, contract or hire developers and maintainers to engineer contributions and mentor internal engineers}.
To build influence organically by on-ramping new developers into a community can be time-consuming, why a firm may consider hiring existing maintainers and developers in leadership positions. As explained by I3, \textit{``If your willing to do a longer-term play, then you get people already in your development team starting to make upstream contributions, then it may take a year or two years depending on what kind of community it is, to have the influence long term. But if you needed that influence yesterday, the only way is to hire someone that is a very strong contributor or maintainer''}. I10 adds, \textit{``We like to hire people in leadership positions. And once we get to two or three people that are in those sort of positions, then we can get started introducing some junior developers''}. This kind of mentoring is further endorsed by I12, \textit{``I would hire the contractor to teach how to do the work. So it's kind of on-the-job-training''}

\textbf{EP4 - Contribute to the development of the OSS project through internal engineers}.
Long-term and sustainable influence is built by directly contributing to the development of the OSS project. These contributions are not limited to code, but may also include \textit{`` \ldots writing documentation, testing, answering questions, doing the mud work, doing a lot of the things that no one wants to do''}, as explained by I3. Developers need to be enabled to actively engage in the community development process without being hindered by internal contribution processes. I17 adds, \textit{``Principally all open source communities are run as meritocracies so we need to be active. If we want to be able to change the direction in [OSS project], we need to produce code and plugins to show that we are part of the community''}.

\textbf{EP5 - Offer the expertize and resources of the firm}.
If a firm holds specific resources, these can also provide valuable contributions to the community. These resources can, for example, be infrastructure-related, but also include soft factors. I1 exemplifies how they provide large-scale testing capabilities, as well as credibility to an OSS project that they run in production, \textit{``if you have big companies like us using [OSS project], it says that it is a viable product''}. I9 adds another example where they provide server space for the community to run compute, test and build processes, enabling the active development in the OSS community.

\textbf{EP6 - Have an active on-line and off-line community presence}.
Community discussions regarding the development of an OSS project 
take place in on-line mediums such as issue-trackers, chats and social media, but also off-line at events and social gatherings such as meetups, conferences, and hackathons. For a firm to grow and leverage its influence, it needs to be present and take an active part in these discussions, and also help to facilitate them, e.g., by arranging their own events. I17 exemplifies, \textit{``Concerning [OSS project], we are extremely active at hackathons... We travel a lot to get and know the people... Recently we hosted a hackathon where we gathered basically all maintainers of the project''}.

These activities should be coordinated internally as highlighted by I1, \textit{``[The Community Manager] ran community activities internally and externally, created awareness of what we were doing in [OSS community], making sure that people contributed to the right projects, submitted abstracts to the right projects, were elected to the right bodies, showed up at the right conferences''}. Having dedicated developer advocates and community managers was a generally recommended practice. This person should be able to mediate and be a spokesperson both of the community and the firm.

\textbf{EP7 - Be open and humble to the OSS community}.
When joining a community, I9 explains, a firm should adapt to the culture and way of working in the community. They should \textit{`` \ldots come in humbly and offer to help in things where they have expertize as opposed to 'We need to do a thing, it's gotta be done this way', then you are going to get an immune reaction if you start that way''}. I12 gives the comparison, \textit{``Joining a new open source community is like moving into a new neighborhood. There is a way of doing things, some of these things are going to be built up during time, and there is going to be inertia. So there are things that are obviously better, that people are going to agree is obviously better, but they are used to the way things are done. So you kind of have to figure out how to bring change gradually. And at the same time is that you figure out how things work. And so, first figure out how the community works before you come in and propose a lot of changes, so don't be excessively critical of the way things are done in a specific neighborhood, then no one is going to listen to you when you propose changes''}. Furthermore, the firm should be \textit{`` \ldots transparent and open about the intentions and the agenda with the community and project, e.g., road-map, what you are keeping closed and for what reason''} as highlighted by I7. This is further emphasized by I13, \textit{``You need to be open and completely honest about the why even if it's for profit because otherwise you just look suspicious''}. The firm should hence differentiate between the communication they use towards the community and that which they use towards for example their employees and customers. 

% Don't expect free contributions without any work, or to get features in on the fly

\textbf{EP8 - Build an inner source culture and practice inside the firm}.
By introducing inner source culture and development practices internally, a firm can help its developers to learn better how to work with external OSS communities, and simplify on-ramps. I13 explains the importance of teaching internal engineers about OSS development practices, such as working distributed and decentralized, \textit{``A lot of companies where everyone sits in the same office all talking among themselves and have meetings just them, and they need to learn how to do everything online, and include the people who are not there''}. In effect, this can create more contributors for the firm and in a longer perspective help raise its influence in OSS communities in general. 

%==============================================
%==============================================
\section{Framework application example: Jenkins and Gerrit}
\label{sec:CaseExample}
%==============================================
%==============================================

In this section, we illustrate through a fictitious example of how the CSF could be applied (cf. descriptive validation~\cite{hevner2004design}), based on a previously studied case which describes how Sony Mobile and its Tools department evolved in their engagement with the communities of the two OSS projects, Jenkins, and Gerrit~\cite{munir2017open}.

Initially, Sony Mobile had a restrictive view of what they shared with the two communities and how they engaged. They focused mainly on doing bug-fixes, general knowledge-sharing and had a community presence limited to online channels, such as mailing lists and issue trackers. The engineers in the Tools department focused on internal work and tailoring of the two OSS projects to internal needs. They further saw that they could create a competitive advantage by keeping internally developed features closed. However, with time, the attitude towards the two communities evolved into a more symbiotic relationship. Sony Mobile and the engineers at the Tools department saw increased benefits with having an active engagement and being more open. As then highlighted by Sony Mobile's Director of OSS operations (I5) - \textit{``\ldots not only should [the tool-chain] be based on OSS, but we should behave like an active committer in the ways we can control, understand and even steer it up to the way we want to have it''}. It is in this context that the aspects of the CSF are analyzed and discussed for Jenkins and Gerrit. 

% The practices used by the Tools department after the shift in how they engaged in the two communities (as reported in earlier work~\cite{munir2017open}) are in line with the practices defined in the CSF and discussed accordingly.

\subsection{Defining the need for influence in the Jenkins and Gerrit communities}

Below we investigate the need for influence in the Jenkins and Gerrit communities by considering the business, technical and community aspects of the CSF from Sony Mobile's point of view.

\subsubsection{Business Aspects}

The connection between the two OSS projects and the business model of Sony Mobile (\textbf{BA1}) was indirect in the sense that the projects were used in the development infrastructure that engineers leverage in the product development inside Sony Mobile. Perceived benefits from the use of the two OSS projects include improved quality of Sony Mobile's end-products, as well as shorter time-to-release and market. Both OSS projects were seen as a commodity and a non-competitive advantage. There were alternative solutions available, but most were proprietary, and the primary motivation for an OSS option was that Sony Mobile could customize the OSS projects based on internal needs much more easily.

The adoption of Jenkins and Gerrit was part of a broader strategy (\textbf{BA2}) of moving Sony Mobile more towards usage of OSS, as well as the adoption of the same tool-chain used by Google in the Android development.

Both communities made up important pools for finding and attracting new and talented employees that could help in adapting the two OSS projects to the preference of Sony Mobile (\textbf{B3}). However, as neither Jenkins or Gerrit was a part of the product or any marketing, there was no need to establish a certain level of visibility or credibility towards the customers (\textbf{B4}).

\subsubsection{Technical Aspects}

% Although the introduction of Jenkins and Gerrit was driven by a top-down initiative, it was also pushed for bottom-up by the engineers. There was a satisfaction among developers in working with the technologies, both in the Tools and Product development departments (\textbf{TA1}).

Both Jenkins and Gerrit made up pivotal parts of the continuous integration tool-chain inside Sony Mobile. Therefore, there were many interactions and dependencies between the two OSS projects and as well as to other tools in the tool-chain (\textbf{TA1}). They had been tailored to internal requirements and supported the development process defined internally.

Sony Mobile was dependent on a stable and secure infrastructure, why they did not need to use the latest or experimental releases (\textbf{TA3}). In general, however, there was an expressed goal to avoid too many patches, and adaptations as the Tools department was limited in resources and had to rely on the community for much of the development (\textbf{TA4}). Further, there was a need to introduce a heavier focus on scalability in the two OSS projects, as they at the time were not optimal in large-scale setups as that used by Sony Mobile (\textbf{TA2}).

\subsubsection{Community Aspects}

Both the Jenkins and Gerrit communities had several firms involved, including direct competitors to Sony Mobile. However, as both OSS projects were seen as non-competitive by Sony Mobile, this presence was not considered as an issue. Few of the existing stakeholder had the equivalent or larger size of installations, which made Sony mobile somewhat unique in its need for improved scalability (\textbf{CA1}).

In general, both communities were very active and diverse concerning contributors and users (\textbf{CA2}). Also, the culture and governance structure was very open for new contributors to join in discussions and rise in rank (\textbf{CA2}). Due to the healthy activity, and meritocratic culture and governance of the communities, it was also deemed easy to increase influence, both organically by introducing employees, but also through hiring new talent as both communities are community-managed(\textbf{CA4}).

\subsubsection{Summary and Goals for Engagement}
% From the business perspective, both Jenkins and Gerrit played an indirect but essential role in improving quality and shortening time-to-release and market of Sony Mobile's handset devices. Due to this loose coupling to the value proposition and revenue streams, the two OSS projects were no longer seen as a competitive advantage, and more as commodity rather than something differentiating. From the technical perspective, there was a heavy reliance and dependency on the projects, both in regards to interactions into the rest of the tool-chain, but also into the related development processes used. Further, there was a defined need to align internal release planning and road-mapping with those of the communities, primarily due to a lack of scalability in the two OSS projects. Finally, from the community perspective, both communities had active and diverse stakeholder populations with an open meritocratic culture and transparent governance structures, which allowed for the potential to gain influence if needed.

Even though classified as non-competitive, there was a defined need to be able to influence the road-maps of the OSS projects, and be able to contribute larger features (e.g., related to improving scalability). Due to the limited size of the Tools department, there was also an expressed goal to be able to find and create collaborations when possible, even with competitors.

\subsection{Defining the engagement activities in the Jenkins and Gerrit communities}

Based on the determined need for influence and goals that were defined, Sony Mobile and its Tools department became more active and open in their engagement with the two communities.

No foundations were surrounding the two projects, why there was no need to attain a specific membership or sponsorship (\textbf{EP1}). However, there were committer groups (\textbf{EP2}), i.e., central parts in the communities' governance~\cite{nakakoji2002evolution}, that Sony Mobile wanted to join. 

Sony Mobile did not see a need to rush and hire engineers from the communities directly (\textbf{EP3}). Instead, they grew their influence organically by introducing their engineers to the communities to the point where they managed to get to positions in the Gerrit committer group. The engineers were given frame agreements for the two communities where they were allowed to contribute freely, both in regards to features and bug-fixes (\textbf{EP4}). With their large set-up and testing infrastructure, Sony Mobile could also contribute to improving the quality of the two tools (\textbf{EP5}).

The engineers at the Tools department were active and visible in both online and offline communication channels (\textbf{EP6}). Their online presence included active participation in discussion and knowledge sharing through mailing lists, issue trackers, chat channels, and webinars. Offline presence included attending conferences, meet-ups, and hackathons. The latter was seen as an essential forum to do quick implementations (cf. just-in-time RE~\cite{ernst2012case}) as often many of the more influential persons in the communities were gathered in the same room. Alongside this active engagement, Sony Mobile had an open attitude towards the community and was transparent with its agenda. Engineers presented how the two projects were setup internally, as well as best practices and know problems when possible. They even talked to and engaged in knowledge-sharing with direct competitors (\textbf{EP7}). 

The engagement and internal development of Jenkins and Gerrit were further seen as a seed to create an inner source initiative inside Sony Mobile, with the ambition to spread into others corners of the Sony Corporation (\textbf{EP8}). The goal was to grow more contributors and active users to Jenkins and Gerrit, but also in other projects, and maybe even create new ones where motivated.

%==============================================
%==============================================
\section{Discussion}
\label{sec:Discussion}
%==============================================
%==============================================

Below we discuss the validity of the CSF and contrast it to related work.

\subsection{Determining the need for Influence in OSS communities}

% Determining which OSS communities that are of strategic importance is not an easy task~\cite{dahlander2008firms}. Three main perspectives on this difficult assessment emerged during the design of CSF and are captured by the proposed business, technical and community aspects.

From a business perspective, as highlighted by I11, the \textit{`` \ldots sticky point is understanding how the value of the open source matches to the value of the business [a firm is] trying to build''}. In this sense, the business model concept provides a useful lens to frame how the OSS helps to create, deliver, and capture value for a firm~\cite{teece2010business}, and more specifically, through the OSS project's connection to the value proposition and revenue streams as pointed out in BA1 of the CSF. As indicated by the diversity of the firms that the interviewees represent, this connection can be made in several ways, as is reported in literature~\cite{sharivar2018business, chesbrough2007open, riehle2010economic, munir2017open}. This is also true on the business strategy level where the firm chooses and configures its business model to compete in its business environment~\cite{dasilva2014business}. Creating or supporting a competing standard, or commoditizing a technology or market are two ways in how a firm can disrupt their competition and pave the way for their own business model, both reported on in the CSF (BA2) and in literature~\cite{west2006challenges}. This alignment between the CSF and literature is further repeated in regards to the importance of an OSS community as a pool for recruitment (BA3), as well as a marketing tool towards customers (BA4)~\cite{henkel2006selective, dahlander2008firms}.

On an implementation level, it is also important to understand the reflections between how the OSS project is used in the internal development and it's strategic importance to a firm~\cite{butler2018investigation, schaarschmidt2015firms}. In the example of Sony Mobile and the communities of Jenkins and Gerrit (see Section~\ref{sec:CaseExample} and~\cite{munir2017open}), the two OSS projects played a less direct part in the firm's value proposition, but a much more significant from a technical perspective. They constituted core parts in the internal development infrastructure (TA1), and the communities were key partners to adapt and maintain the software (TA4). As the fitness-of-use and road-map alignment were not satisfactory, a high level of influence was required (TA2). If the case was otherwise, and the direction being predictable, the need for an active community presence may be less urgent~\cite{butler2018investigation}.

Supportive evidence and alignment can hence be found between literature and many of the aspects identified in the interviews and presented in the CSF. However, aspects that need consideration may be different depending on the firm and community. For example, Sony Mobile had in regards to their customers, no need to prove credibility or visibility in the Jenkins and Gerrit communities, as these OSS projects were used internally and not part of any marketing or key selling points~\cite{munir2017open}. Hence, the business aspect BA4 is not relevant in this case, while in other cases it may.

Other aspects though can be considered more general such as business aspect BA1. An OSS project can, depending on the case, have a more direct or indirect connection with the value proposition and revenue streams of a firms business model. For Red Hat, the connection may most often be direct as they base many of their products on them~\cite{chesbrough2007open}. Conversely, returning to the example of Sony Mobile, Jenkins and Gerrit had a more indirect connection as they enabled a customized development process. As reported, Sony Mobile experienced these community engagements as having a positive impact on time-to-market and quality of their products~\cite{munir2017open}.

% <Community aspects>

% Continue on reaping the benefits, refer to Henkel (2006) and Hassan's survey?

% The benefits of using OSS are well-known, but to reap these benefits it may be necessary have an active engagement in the community and be able to affect the community's RE process of the concerned OSS project. 

% CSF was first drafted based on ten exploratory interviews 

% interviewees from the problem investigation phase

% Use of CSF when joining an existent community vs. creating a new one?

% Why is the project important? Is it part of our product or is it just an infrastructure project? Do we need it to hire talent?...

% Enough with having the community understanding the company's use cases? Enough to be on the End-user committee/being involved in community discussions?

% Risk mitigation, can focus on other things, see the community as an outside vendor/R\&D pool

% Do you depend on it?

% Product companies (e.g., Red Hat) vs. End-user companies (e.g., Comcast)

% Fitness-of-use - Difference between what it does and what you need it to do?

% Visibility/credibility towards customers maybe only is important for certain companies?
% E.g., IBM and Linux in the 2000s

% Business Aspects and Technical Aspects - Box analogy from JB21

% Standardization/commoditization -> substitute and complementary goods

\subsection{Influencing the Requirements Engineering Process in OSS communities}
\label{subsec:disc:influenceing}
As reported in the literature (see Section~\ref{sec:RelatedWork}), the type of governance in OSS communities can vary~\cite{de2007governance, de2013evolution, shaikh2017governing, capra2008framework}. There is also variation in the possibilities and ways how firms can gain influence on the RE processes in a community. 

Among the cases researched, meritocracy seems to be among the more common authority structures~\cite{butler2018investigation, scacchi2010collaboration, mahoney2005non, nakakoji2002evolution, fielding1999shared}. In a meritocratic community, influence on the RE process is gained by proving merit, and as highlighted in the literature, this does not have to be limited to technical contributions~\cite{eckhardt2014merits, o2007emergence}. In essence, it is about earning the trust and status among one's peers in a community~\cite{dahlander2006man, nguyen2017coopetition, nguyen2018do}. Considering the differentiation by De Noni et al.~\cite{de2013evolution} between open-source based or collective communities, this characteristic can be assigned to both, even though the former is described as institutionalized and democratic, and the later as collective and meritocratic. In a purely democratic community, an individual still needs to earn trust, respect, and recognition among its peers to gain responsibilities and authority. In Apache communities, for example, both democratic and meritocratic traits can be found as individuals are voted into leadership positions even though Apache communities are profiled mainly as meritocracies~\cite{fielding1999shared}. Further, as De Noni et al. classifies Apache communities as open-source based, rather than collective, one can view the two categories of authority structures as closely related, as is the way in how influence can be gained on their communities' RE processes.

In communities with a centralized and autocratic authority structure~\cite{de2007governance}, i.e., firm-sponsored or tolerant dictator-based~\cite{de2013evolution}, project leadership is often centered to a single (or limited number of) firm(s) (e.g., Android Open Source Project) or person(s) (e.g., the Linux kernel project). In firm-sponsored communities~\cite{o2007governance, capra2008framework}, specifically those centered around a single firm~\cite{schaarschmidt2015firms, riehle2012single}, where the focus is more on transparency than accessibility~\cite{west2008role}, communities may often be viewed more as user communities and a less open type of software ecosystem~\cite{gawer2014bridging, jansen2012shades}. To gain influence in these types of communities, firms may focus more on direct business relationships (cf.~\cite{baars2012framework, valencca2017theory}). However, this does not prevent meritocratic governance aspects to be present why the community engagement practices as proposed by the CSF may still be relevant. In tolerant dictator-based communities, as shown by Shaikh and Henfridsson~\cite{shaikh2017governing}, there can still be mixes of meritocracy and democracy implemented through different coordination processes. Even if such coordination practices would not be present in an autocratic community, there is still some possibility to influence by earning trust and respect in the community. If a firm can create enough traction among their peers in the community, the project leadership will commonly consider it~\cite{german2003gnome, laurent2009lessons, noll2007innovation, munir2017open}. If not, and an opposing will is strong, part of the community may in worst case move to create their own fork of the project~\cite{nyman2013code}, as was the case with OpenOffice and LibreOffice~\cite{gamalielsson2014sustainability}.

Findings from the interviews regarding engagement practices align with Dahlander and Magnusson~\cite{dahlander2005relationships}, in that influence in a meritocratic OSS community is built through creating a symbiotic relationship with the community. Trust and status are gained through active involvement and respecting its norms and values~\cite{syeed2017measuring, dahlander2005relationships, aagerfalk2008outsourcing, nguyen2018do}. As pointed out by S9, gaining influence may be done through different types of engagements and with varying types of resources, \textit{``It's bringing code, bringing people into influence, in most projects, buying influence is not as easy to do, but you can still spend sponsorship money and port money to make sure that a project is happier or healthier for example''}. I.e., influence may, for example, be gained through providing code contributions as well as more general resources, including financial, aligning with practices reported in literature~\cite{stam2009when, o2007emergence, munir2017open, butler2018investigation, henkel2008champions, dahlander2006man, nguyen2017coopetition, nguyen2018do, omahony2008bounary}. When comparing the communities inside the Linux Foundation and the Apache Software Foundation, S9 describes it as, \textit{`` \ldots you need to show contribution, activity, commitment, leadership, and then you grow through contributions that you take in both foundations''}. Hence, the engagement practices in the CSF are primarily intended for OSS communities where there is a presence of meritocratic coordination processes~\cite{shaikh2017governing}. 

\section{Threats to Validity}
\label{sec:ThreatsToValidity}
%==============================================
%==============================================

% \subsection{Internal Validity}
% When investigating how one factor affects another there may be a risk of further factors affecting the results~\cite{runeson2012casestudy}.

% \subsection{External Validity}
% External validity concerns to what extent the results from a study are generalizable outside of the context of the study~\cite{runeson2012casestudy}.

% \subsection{Construct Validity}
% Construct validity concerns if the factors investigated in a study corresponds to what was originally intended by the researcher~\cite{runeson2012casestudy}.

% \subsection{Reliability}
% Reliability concerns if the produced research results can be replicated or dependent on the researcher~\cite{runeson2012casestudy}.

As presented in Section~\ref{sec:CommunityStrategyFramework}, the CSF covers a broad spectrum of aspects, some more general and applicable than others. One reason for this may be that the 18 interviewees each have extensive personal experience in the field but with different backgrounds, e.g., business or developer-oriented. Another reason may be that they represent 12 different firms (see Table~\ref{tbl:Interviewees}) which in turn may have different use cases and needs. From an external validity perspective~\cite{runeson2012casestudy}, this is positive and indicates a potential of transferability to other firms who are engaged (or are aspiring to) in meritocratic OSS communities. However, as the CSF is based on qualitative data from a limited set of interviewees, quantitative conclusions on generalization will require further validation using statistics based on a population of real-world firms and OSS communities.

Another area regarding external validity is to what extent the practices presented by the CSF actually leads to a gain in influence, and in what contexts. As discussed in Section~\ref{subsec:disc:influenceing}, we believe that there has to be meritocratic coordination processes present~\cite{shaikh2017governing} as the engagement practices proposed in the CSF present make up different ways in how a firm can contribute to and engage with a community to build a symbiotic relationship based on trust, respect, and recognition among its peers. Interviews from the case validation (see Section~ \ref{subsubsec:caseValidation}) supports these arguments as the communities that CaseOrg is engaged in, and in the context which CSF was discussed, were all community-managed and meritocratic. However, external validity is still a limitation in regards to CSF why further empirical validation is needed in future studies, e.g., through the use of case studies and cross-case synthesis.

% Aligning with literature~\cite{dahlander2005relationships, dahlander2006man, munir2017open}, interview findings point to that firms should engage in several different ways and contribute to the development and health of the community. As further discussed in the aforementioned section, this may, however, have limited validity regarding what governance a community has. 
% In firm-sponsored communities specifically~\cite{schaarschmidt2015firms, west2008role}, influence may be more related to building a business relationship with the sponsoring firm. We believe that the practices proposed are mainly valid in contexts where the OSS community, or a legal entity representing it, owns and manages its OSS project~\cite{capra2008framework, o2007governance}. 
% We believe that there has to be meritocratic coordination processes present~\cite{shaikh2017governing} as the engagement practices proposed in the CSF present ways to engage and support an OSS community with the intention to earn merit, trust, respect, and recognition among the peers in the community. 

Regarding the completeness on the aspects and practices presented in the CSF, we again acknowledge that CSF is based on qualitative data from a limited set of interviewees. When performing the interviews in the Interview Validation step (see Section~\ref{subsubsec:InterviewValidation}) of the validation phase we did reach a point of saturation where we observed a tendency of maturity in terms of declining number of emerged codes. This observation was further supported in the Case Validation (see Section~\ref{subsubsec:caseValidation}) as no aspects or practices were added or removed, only further refined. This may point to some level of completeness. However, as presented in Section~\ref{sec:RelatedWork}, there are numerous variations in the characteristics of OSS communities, e.g., in regards to governance structure, demographics, and RE process. Hence, further research and design cycles are needed to validate the CSF and to improve its level of completeness.

\section{Conclusions}
\label{sec:Conclusions}
%==============================================
%==============================================

The focus of this study has been to identify what aspects that firms should consider when they assess their need of influencing the RE process in a meritocratic OSS community (\textbf{RQ1}), as well as what practices that should be considered in order to gain this influence (\textbf{RQ2}). To address these questions we used a design science approach~\cite{wieringa2014design, hevner2004design}. We developed a questionnaire used in ten semi-structured interviews with industry professionals. Inductive coding of interview transcripts~\cite{runeson2012casestudy}, an initial version of a Contribution Strategy Framework was developed. The framework was then validated and refined through seven new interviews and by applying it on a fictitious example of an earlier reported study~\cite{munir2017open}. Finally, a case validation was performed by interviewing four industry professionals from a software-intensive firm engaged in multiple OSS communities. Questions focused on the validity of CSF in the context of the firm's community engagements. In total, 21 interviews were conducted with 18 industry professionals from 12 different software-intensive firms.

The framework consists of aspects and engagement practices. The aspects address \textbf{RQ1} and are divided into business, technical, and community aspects. The two former may be considered to help determine how important an OSS project and its community is from the business and technical perspectives, while the community aspects add the perspective of feasibility and potential in gaining influence in a community. The engagement practices address \textbf{RQ2} and should be seen as a tool-box of ways in how a firm can engage with a community to build influence needed in the community. 
% These characteristics are, according to the interviewees in this study, which leads to influence on the requirements in an OSS community with meritocratic governance, or aspects thereof. 

% The engagement practices are primarily intended for OSS communities where the OSS project is owned and managed by the community or a legal entity (e.g., a foundation) representing the community~\cite{capra2008framework, o2007governance}, and where there are meritocratic coordination processes present~\cite{shaikh2017governing}. In other cases, such as firm-sponsored communities centered around a single firm~\cite{west2008role, schaarschmidt2015firms}, influence may be more related to building a business relationship with the sponsoring firm. This is an example of when it is essential to consider the community aspects when investigating a community, and putting efforts in understanding the community's specific governance, norms, and processes.

As this study uses a qualitative survey approach with a limited sampling of interviewees, further research is needed to validate the CSF through case studies and additional empirical work, both qualitative and quantitative. Along with such research, more theory-grounding work should be performed to further formalize the concept of influence in OSS communities, and how it can be gained, as exemplified by the CSF. Inspiration may be gathered from Valen{\c{c}}a and Alves~\cite{valencca2017theory} in how they generated a theory of power for emerging software ecosystems formed by small-to-medium sized firms.

\section*{Acknowledgments} 
The authors would like to thank the anonymous interviewees for lending their time and expertise, as well as the anonymous reviewers for their valuable feedback. 
This work was funded by the Swedish National Science Foundation Framework Grant for Strategic Research in Information and Communication Technology, project Synergies (Synthesis of a Software Engineering Framework for Open Innovation through Empirical Research), grant 621-2012-5354, and the industrial excellence center EASE (Embedded Applications Software Engineering)\footnote{http://ease.cs.lth.se}.

% \clearpage %% CHECK PAGINATION

%==============================================
%==============================================
\section*{Appendix A - Interview questionnaire}
\label{sec:Appendix}
%==============================================
%==============================================

\begin{itemize}
    \item Do you, in any way, consider or plan how you engage with a community, and where you spend your resource, and to what extent? If yes, how? Is it formalized in any way? How could this be improved/otherwise done
    \item In what ways can you contribute to an OSS community (code, knowledge, socializing, sponsorship)? What roles would you say are involved in these contributions?
    \item How can you gain the power to change or affect (influence) a community in terms of what features gets implemented, and how they are prioritized? (Short- and long-term)
    \item Do you see any connection or consideration between how you engage and invest in a community and the level of influence you need to have in it? How would you describe it?
    \item How can you consider how an OSS and its community creates value for your company? Is there any relation or consequence between this value and how you engage and invest in the community, and the influence you need in it?
    \item Are you aware of the product planning and development in the OSS communities you are involved in? Are your internal product planning and development aligned with the OSS communities'? Do you consider possible dependencies? Do you see a need for it? How would it affect how you engage and invest in the community and the influence you need in it?
    \item Do you consider the motivation and underlying drivers for why you engage and spend your resources in a community? If yes, how? Do these align with how the community creates value for you and its role in your product strategy? Do they align with what you contribute to the communities? What do you see as the main drivers of your company?
\end{itemize}

% \clearpage %% check pagination 

\section*{References}

\bibliography{main}

\end{document}